\documentclass[letterpaper, 10 pt, conference]{ieeeconf}  

\usepackage{graphicx} 
\usepackage{caption}  
\usepackage{float}
\usepackage{amsfonts}
\usepackage{xcolor}
\usepackage{array}
\usepackage{multirow} 
\usepackage{booktabs} 
\usepackage{amsmath} 

\IEEEoverridecommandlockouts                              

\title{\LARGE \bf
NeuroMoE: A Transformer-Based Mixture-of-Experts Framework for Multi-Modal Neurological Disorder Classification
}

\author{Wajih Hassan Raza$^{1}$, Aamir Bader Shah$^{1}$, Yu Wen$^{1}$, Yidan Shen$^{1}$, Juan Diego Martinez Lemus$^{3}$, \\ Mya Caryn Schiess$^{3}$, Timothy Michael Ellmore$^{4}$, Renjie Hu$^{2}$, Xin Fu$^{1, \dagger}$
\thanks{© 2025 IEEE.  Personal use of this material is permitted.  Permission from IEEE must be obtained for all other uses, in any current or future media, including reprinting/republishing this material for advertising or promotional purposes, creating new collective works, for resale or redistribution to servers or lists, or reuse of any copyrighted component of this work in other works.}
\thanks{$^{1}$Wajih Hassan Raza, Aamir Bader Shah, Yu Wen, Yidan Shen, and Xin Fu are with the Department of Electrical and Computer Engineering, University of Houston, USA}
\thanks{$^{2}$Renjie Hu is with the Department of Information Science Technology, University of Houston, USA}
\thanks{$^{3}$Juan Diego Martinez Lemus is with the Movement Disorders \& Neurodegenerative Diseases Division, University of Texas Health Science Center at Houston, USA}
\thanks{$^{4}$ Mya Caryn Schiess is with the Department of Neurology at McGovern Medical School, University of Texas Health Science Center at Houston, USA}
\thanks{$^{5}$ Timothy Michael Ellmore is with the Department of Psychology, The City College of New York, USA}
\thanks{$\dagger$ Corresponding Author email: \tt\small xfu8@central.uh.edu}
}

\begin{document}

\maketitle
\thispagestyle{empty}
\pagestyle{empty}

\begin{abstract}
The integration of multi-modal Magnetic Resonance Imaging (MRI) and clinical data holds great promise for enhancing the diagnosis of neurological disorders (NDs) in real-world clinical settings. Deep Learning (DL) has recently emerged as a powerful tool for extracting meaningful patterns from medical data to aid in diagnosis. However, existing DL approaches struggle to effectively leverage multi-modal MRI and clinical data, leading to suboptimal performance.

To address this challenge, we utilize a unique, proprietary multi-modal clinical dataset curated for ND research. Based on this dataset, we propose a novel transformer-based Mixture-of-Experts (MoE) framework for ND classification, leveraging multiple MRI modalities—anatomical (aMRI), Diffusion Tensor Imaging (DTI), and functional (fMRI)—alongside clinical assessments. Our framework employs transformer encoders to capture spatial relationships within volumetric MRI data while utilizing modality-specific experts for targeted feature extraction. A gating mechanism with adaptive fusion dynamically integrates expert outputs, ensuring optimal predictive performance. Comprehensive experiments and comparisons with multiple baselines demonstrate that our multi-modal approach significantly enhances diagnostic accuracy, particularly in distinguishing overlapping disease states. Our framework achieves a validation accuracy of 82.47\%, outperforming baseline methods by over 10\%, highlighting its potential to improve ND diagnosis by applying multi-modal learning to real-world clinical data.
\newline

\indent \textit{Clinical Relevance}— In neurological disorder (ND) research, a critical gap remains in mapping the continuum of disease progression, from prodromal conditions such as idiopathic REM Sleep Behavior Disorder (iRBD) to fully developed disorders like Parkinson’s Disease (PD). A robust multimodal approach that seamlessly integrates diverse data modalities is essential for accurately predicting and understanding this progression. Advancements in this area would provide the foundation for developing models capable of distinguishing overlapping disease states (e.g., PD, dementia with Lewy bodies, and multiple system atrophy) and predicting the transition from presymptomatic conditions like iRBD to established NDs.

\end{abstract}

\section{INTRODUCTION}
Neurological disorders (NDs) are a leading cause of disability and mortality worldwide, affecting millions of people and imposing a growing burden on healthcare systems. Among them, NDs, such as Parkinson’s disease (PD), Alzheimer’s disease (AD), Dementia with Lewy bodies (DLB), and multiple system atrophy (MSA), are characterized by progressive neuronal loss and irreversible functional decline. The prevalence of these disorders is rising rapidly, with PD alone projected to affect 17.5 million people globally by 2040 \cite{c1}. Despite their impact, the diagnosis of NDs remains largely clinical, often relying on symptom-based criteria that detect the disease only after substantial neuronal degeneration has occurred \cite{c2}. For example, PD is diagnosed based on motor symptoms, but by the time of diagnosis, over 50\% of dopaminergic neurons are already lost \cite{c3}. Similarly, DLB and MSA are often diagnosed late, making early intervention challenging \cite{x1}. These limitations highlight the urgent need for early, accurate, and data-driven diagnostic approaches to improve disease detection and management.


Among various early diagnostic approaches, idiopathic REM sleep behavior disorder (iRBD) has been recognized as a strong prodromal marker for NDs. Longitudinal studies suggest that a majority of iRBD patients will develop such disorders within a decade \cite{c5}, making iRBD an important clinical indicator for identifying individuals at risk. However, iRBD alone is insufficient to distinguish which specific ND a patient will develop, as PD, DLB, and MSA share overlapping pathological mechanisms and symptoms in their early stages. Neuroimaging techniques, particularly magnetic resonance imaging (MRI), have been explored as potential tools to detect early brain changes in patients with iRBD to improve early diagnosis and disease stratification \cite{c4}. Recent studies indicate that MRI can reveal subtle structural and functional alterations in the brains of iRBD patients even before clinical symptoms of PD, DLB, or MSA manifest \cite{b3}. These findings suggest that multimodal MRI approaches, including anatomical MRI (aMRI), diffusion tensor imaging (DTI), and functional MRI (fMRI), could help identify biomarkers that predict disease progression and differentiate between these NDs at an early stage \cite{b4}.



Despite the promise of MRI in detecting early brain changes in iRBD patients \cite{b5, d2, d3}, analyzing and integrating this data presents significant challenges. Traditional approaches rely on manual feature extraction, followed by simpler Machine Learning (ML) models (e.g., logistic regression), which often treat MRI features in isolation and fail to capture complex spatial relationships between brain regions \cite{b6}. While DL and end-to-end models have improved automated feature learning, they still struggle with integrating diverse data modalities, such as MRI and clinical features, due to their differing scales and information content \cite{b7, b8, b9}. Recently, the Mixture-of-Experts (MoE)-equipped multi-modal network has emerged as a promising solution by enabling adaptive prioritization of relevant features across modalities, capturing spatial dependencies, and ensuring computational efficiency through sparse activation mechanisms \cite{d1}. This makes MoEs an ideal candidate for improving early diagnostic accuracy and handling the challenges posed by NDs like PD and iRBD. However, despite their potential, the exploration of MoE-based multi-modal approaches for MRI data is still in its early stages.

To address the challenges of multi-modal learning in ND diagnosis, we propose NeuroMoE, a novel transformer-based MoE framework for classifying PD, iRBD, and healthy controls (HC). Our framework is designed with real-world clinical applicability in mind, leveraging a comprehensive multi-modal dataset that includes MRI scans, clinical assessments, and serum biomarkers. Specifically, we first employ modality-specific transformer encoders to extract informative features from MRI data, capturing spatial relationships and global disease manifestations that traditional methods often fail to leverage. Then, we introduce an MoE block, where each expert specializes in processing a specific data modality, ensuring that distinct patterns across different data types are effectively learned. Finally, to enable adaptive feature fusion, we implement a gating mechanism that dynamically assigns expert weights based on clinical features, allowing the model to tailor its predictions to patient-specific characteristics. By integrating diverse patient data, NeuroMoE effectively models the heterogeneous nature of NDs, achieving a validation accuracy of 82.47\% and outperforming baseline methods by over 10\%, demonstrating its potential as a clinically relevant and practical diagnostic tool for ND diagnosis. Our key contributions are as follows:
\vspace{-2pt}
\begin{itemize} 
\item We leverage a unique and comprehensive multi-modal dataset, including MRIs, clinical and serum biomarkers, to ensure real-world applicability in ND diagnosis.

\item By applying a robust pre-processing pipeline, we harmonize and align multi-modal data, ensuring consistency across different data types.

\item To enable effective feature extraction from each MRI modality, we design modality-specific encoders that capture spatial relationships and disease patterns.


\item We further introduce an MoE block with adaptive gating, enabling dynamic weighting of modality-specific experts for optimized decision-making.

\item Extensive experiments validate our NeuroMoE design, achieving a validation accuracy of 82.47\%, and outperforming baseline methods by over 10\%.
\end{itemize}



\section{Related Works}
\subsection{Early Diagnosis of Neurological Disorders}
Early diagnosis of NDs, including PD, iRBD, DLB, and MSA, remains a major challenge in the clinic. Traditional diagnostic methods rely on clinical symptom assessments and molecular biomarkers, often detecting the disease at advanced stages and limiting opportunities for early intervention \cite{r2}. As a prodromal biomarker, iRBD indicates that up to 80\% of patients eventually develop PD, DLB, or MSA \cite{r3}. However, iRBD alone cannot accurately predict disease progression \cite{r4}. MRI plays a crucial role in early diagnosis by detecting subtle brain changes before clinical symptoms emerge. Advanced MRI modalities, such as aMRI, DTI, and fMRI, provide valuable biomarkers for differentiating NDs and monitoring disease progression \cite{r6}. These techniques detect structural atrophy, white matter integrity disruption, and functional connectivity alterations, which often remain undetectable with conventional biomarkers \cite{r7, r8, r9}. MRI is also essential for evaluating treatment efficacy and refining intervention strategies. By providing non-invasive insights into NDs, it supports early detection and improves patient management \cite{r18}. However, current approaches, especially those based on ML, often focus on a single modality, limiting early detection accuracy \cite{x2, x3}. By combining complementary biomarkers across modalities, our method enhances diagnostic accuracy and improves early detection capabilities, reducing misdiagnosis and missed diagnoses.

\vspace{-1pt}
\subsection{Mixture-of-Experts Multi-modal Networks in Healthcare}
In healthcare, multi-modal DL networks integrate diverse data types, including medical imaging, clinical records, and genomic information, to uncover complex patterns and enhance diagnostic accuracy and patient care. For example, combining chest radiographs with electronic health records has improved disease diagnosis, demonstrating the potential of multi-modal approaches in clinical settings \cite{r11}. Similarly, integrating brain MRI with clinical data has enhanced early detection of NDs, while fusing histopathology images with molecular biomarkers has improved cancer subtype classification \cite{u1, r17}. Recently, mixture-of-experts (MoE) architectures have further advanced this integration by employing specialized sub-networks, or ‘experts,’ each trained to process specific data types or tasks \cite{r15}. This approach not only enhances model performance but also improves computational efficiency by activating only the necessary experts during inference. MoE models have effectively combined imaging and clinical data, while also demonstrating robustness to missing data, leading to more accurate disease classification and prognosis \cite{r12, r14}. Nevertheless, the application of MoE networks in neurology remains in its early stages due to the complexity of neurological data, such as structural and dimensional differences between brain imaging and clinical data, as well as the limited availability of large multi-modal datasets for robust model training. In this work, we pioneer the exploration of MoE multimodal transformers for the classification of neurological disorders, offering a comprehensive framework that achieves high accuracy.

%

\begin{table}[tb!]
\centering
\caption{Summary of various biomarkers across NDs.}
\resizebox{\columnwidth}{!}{  
\begin{tabular}{@{}l c c c@{}}
\toprule
\multirow{2}{*}{\textbf{Biomarkers}} & \multicolumn{3}{c}{\textbf{Neurological Disorder (N = 113)}} \\ 
\cmidrule(l){2-4}
 & \textbf{PD (n=41)} & \textbf{RBD (n=44)} & \textbf{HC (n=28)} \\ 
\midrule
\multicolumn{4}{c}{\textbf{\textit{Clinical Biomarkers}}} \\
Disease Duration & 6.11 ± 3.41 & 6.45 ± 3.12 & 0.00 ± 0.00 \\
PSG result & 0 (18), 1 (23) & 0 (2), 1 (42) & 0 (28), 1 (0) \\
UPDRS part III & 29.75 ± 15.17 & 63.90 ± 20.35 & 2.81 ± 3.11 \\ 
TUG & 18.93 ± 14.61 & 21.95 ± 2.28 & 14.23 ± 1.97 \\ 
H\&Y & 1.73 ± 0.72 & 0.00 ± 0.00  &  0.00 ± 0.00 \\
MoCA & 27.75 ± 2.83 & 18.60 ± 7.96 & 28.54 ± 1.97 \\
UPSIT & 21.59 ± 7.24 & 15.40 ± 2.61 & 35.30 ± 4.51 \\
\midrule
\multicolumn{4}{c}{\textbf{\textit{Serum Biomarkers}}} \\
Vitamin D & 35.68 ± 14.39 & 33.38 ± 7.54 & 36.61 ± 15.41 \\
Uric Acid & 4.34 ± 0.99 & 4.50 ± 0.34 & 4.80 ± 1.34 \\
Serum IFN\(\gamma\) levels & 27.74 ± 54.11 & 4.21 ± 4.00 & 33.01 ± 58.63 \\
\midrule
\multicolumn{4}{c}{\textbf{\textit{MRI Biomarkers}}} \\
aMRI & Imaging & Imaging & Imaging \\
DTI & Imaging & Imaging & Imaging \\
fMRI & Imaging & Imaging & Imaging \\
\bottomrule
\end{tabular}
} 
\caption*{\textit{Note:} For categorical variables (e.g., PSG result), the values (0/1) are presented as observed classes, with the number of patients in parentheses. For continuous and discrete variables, values are expressed as mean ± standard deviation.}
\label{tab:biomarkers}
\end{table}

\section{Multi-Modal Dataset for Neurological Disorders: Parkinson’s Disease and iRBD}
To develop machine learning methods with real-world clinical applicability, we utilize a unique, proprietary, and non-public multi-modal dataset curated by the University of Texas Health (UTHealth) Science Center in Houston for the study of PD and idiopathic iRBD. This dataset originates from a prospective longitudinal cohort study involving 113 individuals aged 40 to 70 years who were clinically diagnosed with PD (41 participants), had polysomnographically confirmed iRBD (44 participants), or served as HC (28 participants). Participant recruitment took place between 2009 and 2018 at the Memorial Hermann Hospital and LBJ sleep laboratories, with clinical assessments conducted at the UTHealth Movement Disorders Subspecialty Clinic (UTMOVE). Individuals with secondary RBD were excluded to maintain cohort specificity. The study was approved by the Institutional Review Board of The UTHealth Science Center at Houston (IRB number: HSC-MS-08-0147). All study procedures were conducted in accordance with institutional guidelines and ethical standards for research involving human participants. Written informed consent was obtained from all participants during the screening visit, before the initiation of any study-related procedures.

To ensure diagnostic accuracy, all iRBD and HC participants underwent comprehensive neurological evaluations to rule out dementia and parkinsonism. The dataset includes cross-sectional baseline data encompassing clinical assessments, neuroimaging scans, and serum biomarker measurements. This multi-modal dataset comprehensively represents disease pathology, facilitating in-depth analysis and classification. Table \ref{tab:biomarkers} provides a summary of the included modalities, detailed as follows:

\begin{itemize}
    \item {\textbf{Clinical Biomarkers} includes several standardized scoring scales to assess disease severity and symptoms: the Modified Hoehn \& Yahr Scale (H\&Y) for disease stage, the Unified Parkinson’s Disease Rating Scale (UPDRS) Part III for motor function, the Timed Up and Go (TUG) Test for gait function, the Montreal Cognitive Assessment (MoCA) and the University of Pennsylvania Smell Identification Test (UPSIT) for non-motor symptoms such as cognitive and olfactory function, and Polysomnography (PSG) to diagnose iRBD.}
    \item {\textbf{MRI Biomarkers} includes aMRI (T1-weighted), DTI, and task-based fMRI. The fMRI was performed during a left finger-tapping task. All participants were imaged using a 3T Philips Medical Systems whole-body MR scanner with an 8-channel SENSE head coil. Detailed acquisition protocols are the same as in \cite{ELLMORE2010645}.}
    \item {\textbf{Serum Biomarkers} includes measurements of several biomarkers: vitamin D (associated with NDs), uric acid (implicated in neuroprotection), and interferon-gamma (IFN$\gamma$) (a cytokine involved in immune response).}
\end{itemize}

\section{NeuroMoE Framework}

\subsection{Overview}

\begin{figure}[tb!]
  \centering
  \includegraphics[width=0.45\textwidth]{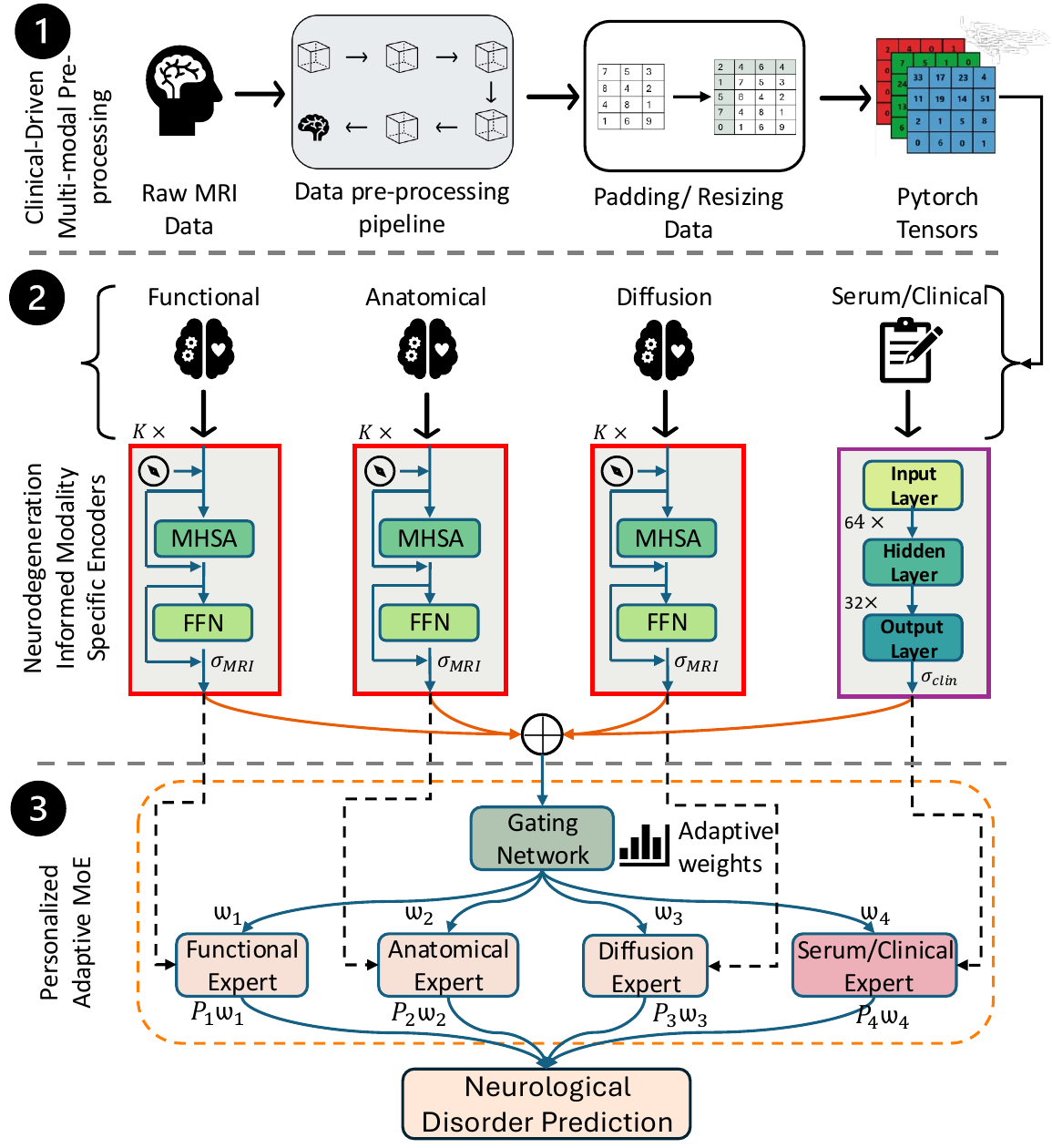}
  \caption{Overall architecture of NeuroMoE.}
  \label{fig:mainarch}
\end{figure}

Although the UT Health dataset provides rich information, effectively integrating diverse multi-modal data remains a significant challenge in the accurate diagnosis of NDs. To address this, we propose NeuroMoE, a unified framework designed to enhance diagnostic accuracy by integrating multi-modal patient data. Specifically, NeuroMoE employs a cross-modal transformer architecture with a specialized MoE module, enabling a more comprehensive understanding of disease pathology. Our framework incorporates three key MRI modalities (aMRI, DTI, and task-based fMRI) along with clinical and serum data, offering a holistic view of NDs to improve clinical decision-making.

As illustrated in Figure 1, the NeuroMoE framework consists of three major stages: \textcircled{1} Clinical-Driven Multi-Modal Pre-processing, \textcircled{2} Neurodegeneration-Informed Modality-Specific Encoder for feature extraction, and \textcircled{3} Personalized Adaptive MoE for final classification. The first stage ensures data consistency, spatial alignment, and noise reduction, optimizing feature representation for downstream learning. The second stage applies our modality-specific encoders, which extracts multi-modality features, and passes them into the MoE block. Finally, the third stage uses a four-expert MoE block, where each expert specializes in processing different feature subsets. The adaptive gating mechanism dynamically integrates expert outputs, capturing complementary modality-specific information. The model then aggregates MoE-generated embeddings and passes them through a classifier to distinguish between PD, iRBD, and HC. By combining these specialized components, NeuroMoE effectively fuses multi-modal data, improving diagnostic precision and facilitating a deeper understanding of disease mechanisms.

\begin{figure}[t]
  \centering
  \includegraphics[scale=0.45]{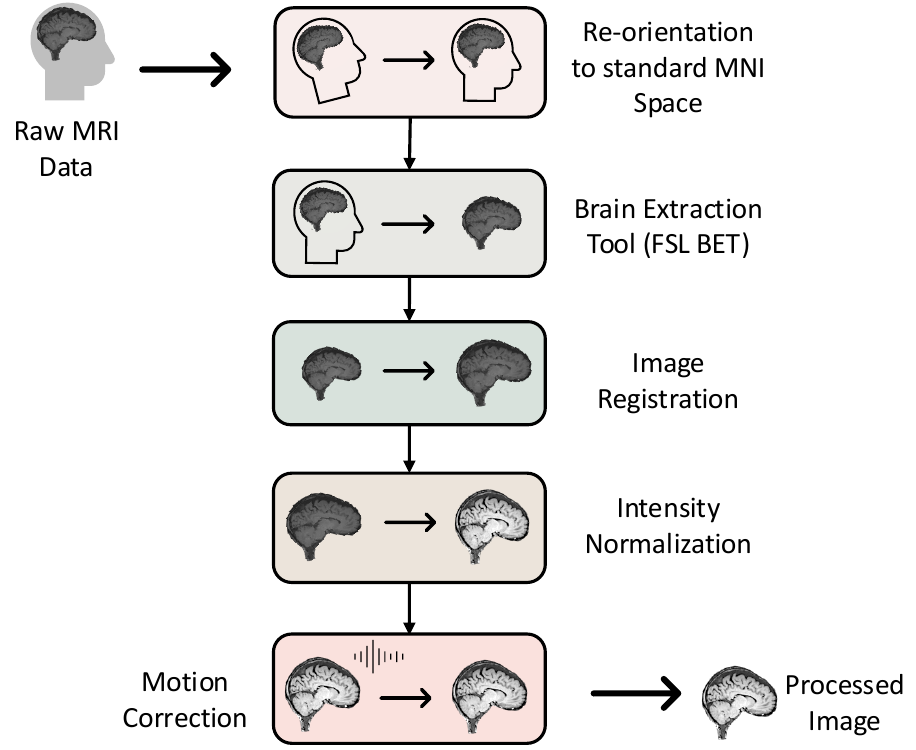}
  \caption{Pre-Processing Pipeline for Multi-Modal MRI.}
  \label{fig:preproc}
\end{figure}


\subsection{Clinical-Driven Multi-Modal Pre-processing}
To ensure the reliability and consistency of multi-modal data, we apply a series of pre-processing steps before model training. MRI data is typically represented as a 3D voxel grid, with each voxel encoding tissue properties at a specific location. Some modalities, such as DTI and fMRI, include a temporal dimension, forming a 4D representation. While providing rich spatiotemporal information, these modalities vary in spatial resolution, intensity distributions, and susceptibility to motion artifacts, making standardization and integration challenging. Addressing these discrepancies is essential for effective multi-modal learning. Therefore, we apply a series of pre-processing steps to harmonize the data before training. Fig. \ref{fig:preproc} illustrates the overall pre-processing pipeline for aligning multi-modal MRI data spatially.

\textbf{MRI Pre-processing:} All MRI images are first reoriented to the Montreal Neurological Institute (MNI-152) space for spatial standardization. Brain extraction is then performed using the Brain Extraction Tool (BET, FSL) to remove non-brain regions. To ensure spatial consistency across subjects and modalities, all MRI volumes are registered to the MNI template using affine and nonlinear transformations (FLIRT \& FNIRT, FSL). For modalities with different intensity distributions, we apply intensity normalization to standardize voxel intensities. Motion correction is also performed to mitigate artifacts and enhance image quality.


\textbf{Serum and Clinical Biomarkers Processing:} To address missing data in 5\% of patients, mean imputation is applied to clinical and serum data. Continuous variables, such as UPDRS scores, are normalized to zero mean and unit variance to ensure consistency in scale for model interpretation. Categorical variables, such as P\&G results, are one-hot encoded to represent each category as binary values, preserving the interpretability of the data.

\subsection{Neurodegeneration-Informed Modality-Specific Encoder}
After pre-processing, all modalities are aligned along the temporal dimension. To effectively process these modalities, we design modality-specific encoders: each MRI modality gets a dedicated encoder to learn different parameters and focus on modality-specific features due to their unique characteristics, while clinical and serum biomarkers share a single encoder due to their similar low-dimensional structure.

\textbf{MRI Encoder:} As each MRI modality has been aligned during pre-processing, we adopted a unified processing pipeline for all MRIs, outlined in the orange box in \textcircled{2} of Fig. \ref{fig:mainarch}. The MRI volumes are first divided into smaller patches of size $4\times4\times4$ voxels. Each patch is flattened and transformed into a feature space using 3D convolutional layers. Given the importance of spatial relationships in NDs, positional embeddings are added to each patch to retain location information. Next, each MRI modality is processed by a transformer encoder with $K$ Multi-Head Self-Attention (MHSA) layers for capturing global spatial relationships, followed by feed-forward networks (FFN) for feature refinement. Finally, the encoded features are pooled via mean aggregation to reduce dimensionality while preserving key information. The output can be represented as:
\begin{equation}
\mathbf{\sigma_{MRI}} = {MP} \Bigg( {E} \Big( \text{MHSA} \big( {PE}(\mathbf{x_{MRI}}) + {P} \big) \Big) \Bigg),
\end{equation}

where $\mathbf{x_{MRI}}$ represents the MRI input data, $P$ represents the positional embeddings, $PE$ represents the patch embeddings, $E$ represents the encoder layer, $MP$ shows the mean pool and $\mathbf{\sigma_{MRI}}$ represent the output of MRI encoder.

\textbf{Serum/Clinical Encoder:} Given the nature of serum and clinical biomarkers, we propose a shared encoder, which uses a lightweight fully-connected neural network (FCN) to process these features, as shown in the purple box of \textcircled{2} in Fig. \ref{fig:mainarch}. The input passes through three full-connected layers. In each hidden layer, we apply ReLU activation functions to introduce non-linearity. The final output is obtained after the last fully connected layer, followed by a linear transformation that produces logits for classification. To mitigate overfitting, we apply a 30\% dropout rate between the layers. This architecture ensures efficient processing of low-dimensional features while enabling effective classification.

\begin{figure}[tb!]
  \centering
  \includegraphics[width=0.42\textwidth]{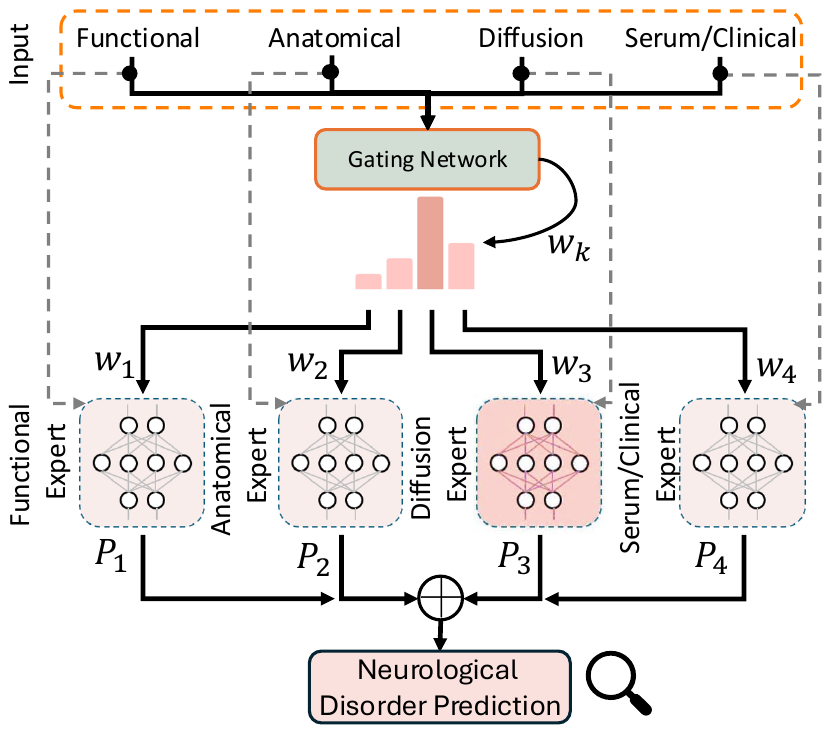}
  \caption{Workflow of expert utilization via dynamic gating.}
  \label{fig:moe}
\end{figure}

\subsection{Personalized Adaptive MoE}
To process features from different modalities effectively, we design modality-specific expert blocks that dynamically adapt to each patient’s unique data, capturing the varying influence of each modality. As shown in Fig. \ref{fig:moe}, each expert network is specialized to process features extracted from its corresponding modality-specific encoder. However, the contribution of each modality can vary depending on the patient’s attributes. To account for this variability, we implement a gating network that dynamically assigns weights to each expert’s output, allowing the model to prioritize the most relevant modalities for each patient.


\textbf{Modality-specific Experts:} Building on the previous design, our framework includes four modality-specific experts: Functional Expert, Anatomical Expert, Diffusion Expert, and Serum/Clinical Expert. Each expert processes the features extracted by the corresponding encoders, capturing the unique aspects of each modality. These experts refine the modality-specific features and produce predictions, which are then weighted and combined by the gating network for final classification. Specifically, each expert is a lightweight FFN with two fully connected layers and ReLU activations, where the output layer has three nodes corresponding to the classes (PD, iRBD, HC). Since each expert is modality-specific, its parameters are learned independently. After processing, each expert generates a class prediction for its modality.

%


\textbf{Gating Network:} To optimally integrate the strengths of all experts, the gating network adaptively determines their contributions by computing a set of weighting coefficients based on clinical features. Rather than selecting specific experts, it assigns a weight vector $\mathbf{\omega} \in \mathbb{R}^4$, where each element represents the contribution of a corresponding expert. The gating network learns to assign weights to each expert by back-propagating classification errors. It dynamically selects experts based on the input features, allowing the model to effectively leverage information from each modality and tailor predictions to the unique characteristics of each subject.

The gating network consists of two hidden layers with \textit{ReLU} and dropout regularization. The output layer applies a \textit{softmax} to produce normalized gating weights. The final prediction $P_{out}$ is a weighted sum of all outputs:
\vspace{-3pt}
\begin{equation}
    P_{out} = \sum_{i=1}^{4} w_i \cdot P_i ,
\end{equation}

where $P_i$ is the prediction from the $i$-th expert, and $w_i$ is its corresponding gating weight. This adaptive weighting mechanism ensures that more relevant experts have a greater impact on the final decision, leveraging complementary modality information for improved classification. To prevent over-reliance on any single expert and promote balanced utilization of all modalities, we introduce a regularization term in the gating network. This term penalizes imbalanced gating weight distribution across experts. Specifically, the regularization term $R_{balance}$ is given by:
\begin{equation} 
R_{balance} = \frac{1}{4} \sum_{k=1}^{4} \left( \frac{1}{N} \sum_{j=1}^{N} w_{k}^{j} - \frac{1}{4} \right)^2, 
\end{equation}

where $w_{k}^{j}$ represents the gating weight assigned to the $k^{th}$  expert for the $j^{th}$ sample in a batch of size $N$.

\section{Evaluation}
\vspace{-3pt}
\subsection{Experiment Settings}
We evaluated our method by implementing it in Pytorch and validating it on the dataset introduced in Section III. Specifically, the dataset was split with 80\% used for training and 20\% reserved for testing. Since our dataset is unique and one of a kind, it offers a realistic assessment of practical performance and complexity not yet available in conventional datasets. As a result, direct comparisons are impractical. Consequently, our evaluation naturally centered on benchmarking against established open-source models. As baselines, we selected the popular Swin Transformer tiny (Swin-T) model, implemented independently for each MRI modality. Additionally, we compared our method with several recent SOTA models that are open-sourced and retrained on our dataset \cite{e1, e2}.


During training, we used an 8-batch size and the Adam optimizer with an initial learning rate of 0.001. A cosine-annealing scheduler was applied for dynamic learning rate adjustments. Due to the smaller batch size, gradient accumulation with a step size of 4 stabilized gradients. The model was trained for 100 epochs, and the best model was saved based on test accuracy. Our architectural hyperparameters were guided by empirical results and established MoE practices. The number of experts (four) matches the modality encoders to ensure specialization. Attention heads and transformer depth were chosen to balance expressiveness with data size, while the two-layer gating network captures clinical-feature-driven weighting without overfitting. These choices were validated qualitatively via training stability and quantitatively through superior model performance.

\begin{table}[t]
\centering
\caption{Comparison of NeuroMoE with baselines.}
\resizebox{\columnwidth}{!}{%
\begin{tabular}{lccc}
\toprule
\textbf{Methods} & \textbf{Modality} & \textbf{Accuracy} & \textbf{F1-score} \\
\midrule
Swin-T & aMRI & 56.25\% & 55.30\% \\
Swin-T & DTI  & 56.82\% & 55.78\% \\
Swin-T & fMRI  & 58.78\% & 57.04\% \\
Dynamic Image \cite{e1} & aMRI & 68.43\% & 66.82\% \\
LiuNet \cite{e2} & aMRI & 70.00\% & 69.70\% \\
\textbf{NeuroMoE} & aMRI, DTI, fMRI, serum/clinical & \textbf{82.47\%} & \textbf{81.25\%} \\
\bottomrule
\end{tabular}
}

\label{tab:b}
\end{table}
\vspace{-3pt}
\subsection{Results Analysis}
\textbf{Diagnosis Accuracy} Table \ref{tab:b} presents the experimental results for each baseline model and our proposed architecture. As shown, our framework consistently outperforms the baseline methods. Specifically, among the single MRI modalities, fMRI achieves the highest performance, followed by aMRI and DTI. This demonstrates the limitations of relying on a single modality for classifying NDs, as each modality provides distinct and complementary information. In contrast, our proposed MoE framework achieves the highest accuracy of 82.47\% (F1: 81.25\%), significantly surpassing the baseline methods. By optimally integrating the strengths of each modality, our framework overcomes the challenges of multi-modal data integration and adapts to diverse input characteristics to enhance accuracy.

\begin{figure}[tb!]
  \centering
  \includegraphics[scale = 0.40]{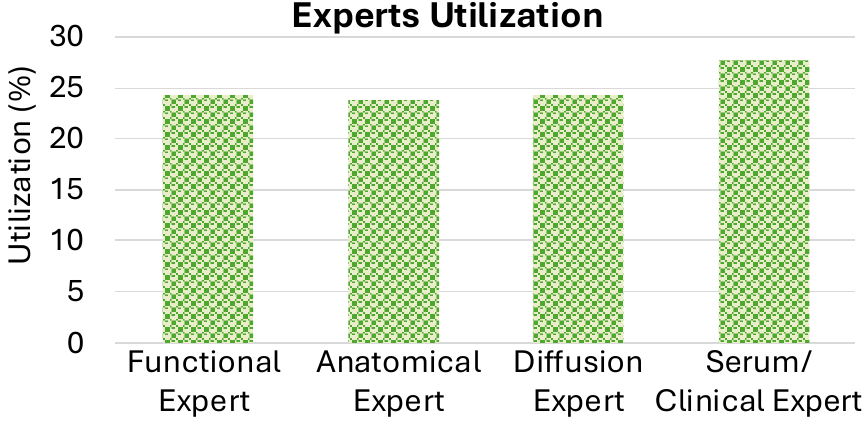} 
  \caption{Average expert utilization with regularization.}
  \label{figure4}
\end{figure}

\textbf{Expert Utilization} Fig. \ref{figure4} shows the average expert utilization within our framework, demonstrating that the MoE model functions as intended, with each modality contributing meaningfully to the predictions. While individual patients may have different modality weightings, the average utilization indicates that all modalities are effectively leveraged. The slight increase in the serum/clinical expert's contribution suggests that clinical features often play a leading role in the prediction process, which is expected since MRI interpretations are typically guided by clinical information for more accurate diagnostic decisions. This balance highlights the complementary nature of multimodal data in enhancing classification performance. Moreover, dynamic gating enhances both accuracy and interpretability by highlighting the most influential experts per patient, allowing clinicians to trace decisions back to key modalities or features.



\textbf{Ablation Study} To validate our model, we conducted a series of ablation experiments by systematically removing each modality and its corresponding encoder from the full model configuration. In each case, we re-evaluated the model with the MoE block and the remaining encoders to quantify the contribution of each modality to overall performance. Additionally, we examined a simplified version of NeuroMoE without the gating mechanism to assess its impact. As shown in Table \ref{tab:c}, the full model achieved the highest accuracy of 82.47\%. Performance declined when individual encoders were removed, with the model without the aMRI encoder achieving 78.08\%, followed by the fMRI encoder (77.73\%), the DTI encoder (77.26\%), and the serum/clinical encoder (75.12\%). The lowest accuracy (65.17\%) was observed when the MoE block operated without gating, indicating that a static combination of experts fails to fully leverage the dataset’s dynamic nature. These results highlight the importance of the gating mechanism and the serum/clinical encoder, as all modalities contribute to optimal accuracy.

\begin{table}[tb!]
\centering
\caption{Ablation results across different configurations.}
\begin{tabular}{lccc}
\toprule
\textbf{Methods} & \textbf{Accuracy} & \textbf{F1-score} \\
\midrule
w/o gate & 65.17\% & 63.49\% \\
w/o serum/clinical encoder & 75.12\% & 73.88\% \\
w/o DTI encoder & 77.26\% & 76.12\% \\
w/o fMRI encoder & 77.73\% & 76.52\% \\
w/o aMRI encoder  & 78.08\% & 76.97\% \\
\textbf{full} & \textbf{82.47\%} & \textbf{81.25\%} \\
\bottomrule
\end{tabular}
\label{tab:c}
\end{table}

\section{Limitation and Future Work}
Due to limitations in the available clinical dataset, our analysis focused on predicting PD and iRBD. However, the framework is easily extendable to other NDs like DLB, MSA, and AD, once corresponding datasets are available. Also, incorporating additional modalities and applying the model to longitudinal studies could provide deeper insights beyond early diagnosis. Another avenue worth exploring is the influence of clinical features on gating weights, which could further assist clinicians in decision-making.


\section{Conclusion}
In this paper, we presented a novel MoE framework for the multi-modal diagnosis of NDs such as PD and iRBD. To ensure real-world applicability, we utilized a unique clinical dataset to guide the design of our framework, enhancing practical diagnostic accuracy. Specifically, we demonstrated that while individual modalities may not achieve optimal performance on their own, our multi-modal approach effectively integrates the strengths of each modality. Moreover, through an adaptive gating mechanism, we showed that classification accuracy could be further improved by dynamically adjusting the importance of each modality based on the input data. Moreover, NeuroMoE is designed for clinical integration, with lightweight experts enabling fast, parallel inference suitable for hospital systems and diagnostic dashboards. Experimental results also showed that our approach achieved a validation accuracy of 82.47\%, surpassing baseline methods by over 10\%, which underscores the effectiveness of multi-modal learning in real-world clinical applications.

\section*{Acknowledgment}
This work is partially supported by NSF grants CCF-2130688 and CNS-2107057. We extend our sincere gratitude to the patients and their families for participating in the trial. We also thank the Adriana Blood Endowment to Dr. Mya Schiess for funding the clinical RBD study.
\bibliographystyle{IEEEtran}
\bibliography{refs.bib}

\end{document}